\documentclass[preprint,prd,nofootinbib,tightenlines,groupedaddress,amsmath,amssymb]{revtex4}

\usepackage{bm}
\usepackage{color}
\usepackage{graphicx}
\usepackage[hypertex]{hyperref}
\usepackage{multirow}

\begin{document}
\baselineskip=15pt \parskip=3pt

\vspace*{3em}

\title{Effects of Family Nonuniversal $\bm{Z'}$ Boson on Leptonic Decays \\
of Higgs and Weak Bosons}

\author{Cheng-Wei Chiang,$^{a,b,c}$ Takaaki Nomura,$^a$ and Jusak Tandean$^d$}
\affiliation{$^a$Department of Physics and Center for Mathematics and Theoretical Physics,
National Central University, Chungli 320, Taiwan
\bigskip \\
$^b$Institute of Physics, Academia Sinica,\\ Taipei 119, Taiwan
\bigskip \\
$^c$Physics Division, National Center for Theoretical Sciences,\\ Hsinchu 300, Taiwan
\bigskip \\
$^d$Department of Physics and Center for Theoretical  Sciences, National Taiwan University,\\
Taipei 106, Taiwan \\
$\vphantom{\bigg|_{\bigg|}^|}$}


\pacs{12.60.-i, 14.70.Pw, 14.80.Bn, 12.60.Cn}

\begin{abstract}
Though not completely a surprise according to the standard model and existing indirect
constraints, the Higgs-like particle, $h$, of mass around 125 GeV recently observed at
the LHC may offer an additional window to physics beyond the standard model.
In particular, its decay pattern can be modified by the existence of new particles.
One of the popular scenarios involves a $Z'$ boson associated with an extra Abelian gauge
group.
In this study, we explore the potential effects of such a boson with family-nonuniversal
couplings on the leptonic decays of~$h$, both flavor-conserving and flavor-changing.
For current constraints, we take into account leptonic decays at the $Z$ pole, LEP II
scattering data, limits on various flavor-changing lepton transitions, and lepton
magnetic dipole moments.   Adopting a model-independent approach and assuming that
the $Z'$ has negligible mixing with the $Z$ boson, we find that present data allow
the $Z'$ effects to reach a~few percent or higher on $h$ decays into a pair of leptons.
Future measurements on $h$ at the LHC or a~linear collider can therefore detect the $Z'$
contributions or impose further constraints on its couplings.  We also consider $Z'$-mediated
four-lepton decays of the $Z$ and $W$ bosons.

\end{abstract}

\maketitle

\section{Introduction}

The recent discovery at the LHC \cite{lhc} of a new particle having mass about 125\,GeV and
other properties compatible with those of the standard model (SM) Higgs boson undoubtedly has
far-reaching implications for efforts to search for new physics beyond the SM.
Particularly, all new-physics models would have to include such a particle, hereafter denoted
by $h$, as one of their ingredients.  In general, different models would have different
production rates and decay patterns for $h$ because of contributions from and/or mixing with
other new particles.  It is therefore important to have a~detailed study of the characteristic
quantum numbers of $h$ and its interactions with known SM particles.
It is hoped that through such an analysis, the newly discovered particle will provide us with
hints of new physics.

One of the possible scenarios for new physics is the existence of an extra U(1) gauge group
involving a massive gauge particle, the $Z'$ boson.  Such a gauge symmetry may have its origin
from various grand unified theories, string-inspired models, dynamical symmetry breaking
models, and little Higgs models~\cite{Nardi:1993ag,Langacker:2008yv}, just to name a few.
The $Z'$ boson in different representative models has been directly searched for at colliders
as well as indirectly probed via a variety of precision
data~\cite{brooijmans-pdg,delAguila:2011yd}, putting limits on its gauge coupling and/or mass.
Generally speaking, the $Z'$ couplings to SM fermions can be family universal or nonuniversal.
The latter case has especially attracted a~lot of interest in recent years due to its many
interesting phenomenological implications~\cite{Yue:2002ja,Chiang:2011cv}.

In this work, we focus on family-nonuniversal interactions of the $Z'$ with the neutrinos and
charged leptons and explore constraints on its relevant couplings from a number of experiments
on transitions involving leptons in the initial and final states, plus possibly a photon.
These processes suffer less from QCD corrections and hadronic uncertainties than processes
involving hadrons.  Moreover, many of such experiments have been performed at a high precision,
imposing relatively stringent constraints on any possible new interactions.

More specifically, we assume that the $Z'$ boson arises from a new U(1) gauge symmetry,
interacts with leptons in a family-nonuniversal way, and has negligible mixing with the $Z$
boson for simplicity, but otherwise adopt a model-independent approach to make the analysis
as general as possible. Due to the family nonuniversality, such a $Z'$ boson would feature
flavor-changing leptonic couplings at tree level, leading to the distinctive signature of
lepton-flavor violation. However, lepton-flavor violating processes have been searched for
at colliders with null results.  We therefore examine a~number of flavor-conserving and
flavor-changing processes to evaluate constraints on the leptonic $Z'$ couplings.
The results are then used to estimate $Z'$ contributions to both flavor-conserving and
flavor-changing decays of the Higgs boson into a pair of leptons, \,$h\to l^+l^{\prime-}$,\,
at the one-loop level. As $h$ will be probed with increasing precision at the LHC in coming
years, and even more so if a~Higgs factory is built in the future, the acquired data could
reveal the signals of the $Z'$ boson which we consider, assuming that $h$ is a SM-like Higgs
boson.
Last but not least, we will compute the $Z'$ effects on several four-lepton decays of the $W$
and $Z$ bosons and make comparison with the data if available.  The LHC may also be sensitive
to such indirect indications of the $Z'$ presence.  The information on the $Z'$ gained from
the $h$, $W$, and $Z$ measurements would be complementary to that from the $Z'$ direct searches.

This paper is organized as follows.
We present the interactions of the $Z'$ boson with the leptons in
Section~\ref{sec:interactions}, allowing for in particular flavor-nonuniversal couplings.
In Section~\ref{sec:constraints}, we study constraints on the couplings of the~$Z'$ from
$Z$-pole data, cross sections of \,$e^+e^-$\, scattering into lepton-antilepton pairs
measured at LEP~II, various experimental limits on low-energy flavor-changing processes
involving charged leptons, and measurements of their anomalous magnetic moments.
In Section~\ref{sec:predictions}, we proceed to make predictions on both flavor-conserving
and flavor-changing decays of $h$ into a pair of charged leptons. We also explore the $Z'$
contributions to $W$ and $Z$ decays into four leptons.  We summarize our findings
in Section~\ref{sec:summary}.

\section{Interactions \label{sec:interactions}}

The Lagrangian describing the interactions of the $Z'$ boson with neutrinos $\nu_j'$
and charged leptons $\ell_j'$ can be expressed as
\begin{eqnarray}
{\cal L} \,\,=\,\, -g_{Lj}'\,\bar\nu_j'\gamma^\lambda P_L^{}\nu_j'\,Z_\lambda' \,-\,
\bar\ell_j'\gamma^\lambda \bigl( g_{Lj}'P_L^{}+g_{Rj}'P_R^{}\bigr)\ell_j'\,Z_\lambda' ~,
\end{eqnarray}
where summation over \,$j=1,2,3$\, is implied, the primes of the lepton fields refer to their
interaction eigenstates, \,$P_{L,R}^{}=\frac{1}{2}(1\mp\gamma_5^{})$,\, and the parameters
$g_{Lj,Rj}'$ are generally different from one another, reflecting
the family nonuniversality.\footnote{Note that throughout the paper, we use $\ell$ to denote
the triplet of charged leptons, \,$(\ell_1,\ell_2,\ell_3)=(e,\mu,\tau)$,\,
and $l$ to refer to an individual charged lepton in general.}
The Hermiticity of $\cal L$ requires these coupling constants to be real.
Since each of the left-handed neutrinos and its charged counterpart form a SM weak doublet,
they share the same~$g_{Lj}'$.
Since we are concerned with processes below the electroweak scale, we do not consider right-handed
neutrinos in the low-energy spectrum.
The $Z'$ may also have couplings to quarks and other nonstandard fermions, but we do not
address them in this analysis.

The interaction states are related to the mass eigenstates $\nu_j^{}$ and $\ell_j^{}$ by
\begin{eqnarray} \label{V}
\nu_{iL}' \,\,=\,\, (V_\nu)_{ij\,}^{}\nu_{jL}^{} ~, \hspace{5ex}
\ell_{iL}' \,\,=\,\, (V_L)_{ij\,}^{}\ell_{jL}^{} ~, \hspace{5ex}
\ell_{iR}' \,\,=\,\, (V_R)_{ij\,}^{}\ell_{jR}^{} ~,
\end{eqnarray}
where \,$f_{L,R}^{}=P_{L,R\,}^{}f$\, for fermion $f$ and the 3$\times$3 matrices
$V_{\nu,L,R}$ are unitary.
In terms of the mass eigenstates, one can then write
\begin{eqnarray}   \label{Ll}
{\cal L} \,\,=\,\, -b_\nu^{ij}\,\bar\nu_i^{}\gamma^\lambda P_L^{}\nu_j^{}\,Z_\lambda' \,-\,
\bar\ell_i^{}\gamma^\lambda\Bigl(b_L^{\ell_i\ell_j}P_L^{} +
b_R^{\ell_i\ell_j}P_R^{}\Bigr)\ell_j^{}\,Z_\lambda' ~,
\end{eqnarray}
where summation over \,$i,j=1,2,3$\, is implied and
\begin{eqnarray}  \label{bV}
b_\nu^{rs} \,\,=\,\, (V_\nu)_{rj}^\dagger\,g_{Lj\,}'(V_\nu)_{js}^{} ~, \hspace{5ex}
b_L^{\ell_r\ell_s} \,\,=\,\, (V_L)_{rj}^\dagger\,g_{Lj\,}'(V_L)_{js}^{} ~, \hspace{5ex}
b_R^{\ell_r\ell_s} \,\,=\,\, (V_R)_{rj}^\dagger\,g_{Rj\,}'(V_R)_{js}^{} ~.
\end{eqnarray}
It follows that
\begin{eqnarray}  \label{bb*}
b_\nu^{rs} \,\,=\,\, \bigl(b_\nu^{sr}\bigr)^* ~, \hspace{5ex}
b_{L,R}^{\ell_r\ell_s} \,\,=\,\, \bigl(b_{L,R}^{\ell_s\ell_r}\bigr)^* ~, \hspace{5ex}
b_\nu^{rs} \,\,=\,\, {\cal U}_{ri}^\dagger\,b_L^{\ell_i\ell_j\,} {\cal U}_{js}^{} ~,
\end{eqnarray}
where \,${\cal U}=V_L^\dagger V_\nu^{}$.\,
Hence family nonuniversality implies that the $Z'$ interactions with the leptons can be
flavor violating at tree level.
Furthermore, $b_\nu^{ij}$ and $b_L^{\ell_i\ell_j}$ are generally unequal.

\section{Constraints\label{sec:constraints}}

\subsection{High-energy observables\label{high}}

We begin with the determination of constraints from the existing data on the $Z$-boson
decays \,$Z\to l^+l^-$\, and~\,$Z\to\nu\bar\nu$.\,
For the former, the amplitude takes the form
\begin{eqnarray} \label{MZ2ll}
{\cal M}_{Z\to l^+l^-} \,\,=\,\, \bar l\gamma_\lambda^{}
\bigl(L_{ll}^{}P_L^{} \,+\, R_{ll}^{}P_R^{}\bigr)l\,\varepsilon_Z^\lambda ~,
\end{eqnarray}
where $L_{ll}$ and $R_{ll}$ contain both SM and $Z'$ contributions and are given by
\begin{eqnarray} & \displaystyle
L_{ll}^{} \,\,=\,\, g_L^{\rm sm} \bigl(1+\epsilon_L^{llZ}\bigr) ~, \hspace{5ex}
R_{ll}^{} \,\,=\,\, g_R^{\rm sm} \bigl(1+\epsilon_R^{llZ}\bigr) ~,
& \\ & \label{gsm} \displaystyle
g_L^{\rm sm} \,\,=\,\,
\frac{g}{2c_{\rm w}^{}}\bigl(2s_{\rm w}^2-1\bigr) ~, \hspace{5ex}
g_R^{\rm sm} \,\,=\,\, \frac{g\,s_{\rm w}^2}{c_{\rm w}^{}} ~, \hspace{5ex}
c_{\rm w}^{} \,\,=\,\, \sqrt{1-s_{\rm w}^2}
\end{eqnarray}
with as usual the weak coupling constant $g$ and \,$s_{\rm w}^2=\sin^2\theta_{\rm W}^{}$\,
involving the Weinberg angle.
In the absence of $Z$-$Z'$ mixing,\footnote{This is a reasonable approximation based on
the findings of various analyses that the mixing parameter typically has an upper bound
inferred from data of a~few times $10^{-2}$ for
\,$m_{Z'}^{}\raisebox{0.4ex}{$\scriptstyle\,\lesssim\,$}100$\,GeV\, or lower for greater $Z'$
masses~\cite{Langacker:2008yv,Chiang:2011cv,Hook:2010tw}.
Moreover, there are scenarios in which $Z$-$Z'$ mass-mixing is absent, because no Higgs bosons
in the theory carry both the electroweak and extra-U(1) quantum numbers, and kinetic mixing
between the hypercharge and extra-U(1) gauge bosons is naturally small~\cite{Carone:1994aa}.}
the $Z'$ effects modify the $Zl^+l^-$ vertex and
leptonic self-energy diagrams at the one-loop level.
We obtain
\begin{eqnarray} \label{epsilonllZ} &&
\epsilon_{\sf C}^{llZ} \,\,=\,\, {\cal F}_Z(\delta)\,
\mbox{\footnotesize$\displaystyle\sum_{f=e,\mu,\tau}$}\bigl|b_{\sf C}^{fl}\bigr|^2 ~, \hspace{5ex}
\delta \,\,=\,\, \frac{m_{Z'}^2}{m_Z^2} ~, \hspace{5ex} {\sf C} \,\,=\,\, L, R ~,
\nonumber \\
{\cal F}_Z(\delta) &=& \frac{1}{16\pi^2} \biggl\{
-\frac{7}{2} - 2\delta - (3+2\delta)\ln\delta -
2(1+\delta)^2\biggl[\ln\delta\;\ln\frac{\delta}{1+\delta}
+ {\rm Li}_2^{}\biggl(-\frac{1}{\delta}\biggr)\biggr]
\nonumber \\ && \hspace*{7ex} -\;
i\pi\, \biggl[ 3+2\delta+ 2(1+\delta)^2\,\ln\frac{\delta}{1+\delta} \biggr] \biggr\} ~,
\end{eqnarray}
where Li$_2$ is the dilogarithm.
The expression for the real part of ${\cal F}_Z$ has been derived previously~\cite{Carone:1994aa}.
The relevant observables here are the forward-backward asymmetry at the $Z$ pole and decay rate
\begin{eqnarray}   \label{GZ2ll}
A_{\rm FB}^{(0,l)} \,\,=\,\, \frac{3}{4}\,A_e^{}A_l^{} ~, \hspace{5ex}
\Gamma_{Z\to l^+l^-} \,\,=\,\, \frac{\sqrt{m_Z^2-4m_l^2}}{16\pi\,m_Z^2}\;
\overline{|{\cal M}_{Z\to l^+l^-}|^2} ~,
\end{eqnarray}
where
\begin{eqnarray}   \label{Al}
A_l^{} \,\,=\,\, \frac{|L_{ll}|^2-|R_{ll}|^2}{|L_{ll}|^2+|R_{ll}|^2} ~, \hspace{5ex}
\overline{|{\cal M}_{Z\to l^+l^-}|^2} \,\,=\,\,
\frac{2}{3}\bigl(|L_{ll}|^2+|R_{ll}|^2\bigr) \bigl(m_Z^2-m_l^2\bigr)
\,+\, 4\,m_l^2\,{\rm Re}\bigl(L_{ll}^*R_{ll}^{}\bigr) ~. ~~~
\end{eqnarray}

In \,$Z\to\nu\bar\nu$,\, the $Z'$-loop contributions are analogous to those
in the charged-lepton case, but without the right-handed couplings.
Since the unobserved neutrinos in the final state may belong to different mass eigenstates,
we can express the amplitude as
\begin{eqnarray} \label{MZ2nn}
{\cal M}_{Z\to\nu_r^{}\bar\nu_s^{}} \,\,=\,\, \frac{g}{2c_{\rm w}^{}}\,
\bar\nu_r^{}\gamma_\lambda^{}N_{rs}^{}P_L^{}\nu_s^{}\,\varepsilon_Z^\lambda ~, \hspace{5ex}
N_{rs}^{} \,\,=\,\, \delta_{rs}^{} \,+\, {\cal F}_Z(\delta)\,
\mbox{\footnotesize$\displaystyle\sum_j$}\, b_\nu^{rj\,}b_\nu^{js} ~.
\end{eqnarray}
Since $b_\nu^{ij}$ and $b_L^{\ell_i\ell_j}$ are related according to Eq.\,(\ref{bb*}),
summing over the final neutrinos then results in the decay rate
\begin{eqnarray} \label{GZ2nn}
\Gamma_{Z\to\nu\bar\nu}^{} \,\,=\,\,
\mbox{\footnotesize$\displaystyle\sum_{\scriptstyle r,s}$}\,\Gamma_{Z\to\nu_r^{}\bar\nu_s^{}}
\,\,=\,\,
\frac{g^2m_Z^{}}{96\pi\,c_{\rm w}^2}\,
\mbox{\footnotesize$\displaystyle\sum_{\scriptstyle r,s}$}\,|N_{rs}|^2
\,\,=\,\,
\frac{g^2m_Z^{}}{96\pi\,c_{\rm w}^2}\,
\mbox{\footnotesize$\displaystyle\sum_{\scriptstyle l=e,\mu,\tau}$}\,
\bigl|1+\epsilon_L^{llZ}\bigr|^2 ~.
\end{eqnarray}
Accordingly, this channel may offer a complementary probe for~$b_L^{\ell_i\ell_j}$.
From the formula for $\epsilon_{\sf C}^{llZ}$ in~Eq.\,(\ref{epsilonllZ}), we can then see
that these $Z$ decay modes are potentially sensitive to not only the flavor-conserving $Z'$
couplings, but also the flavor-changing ones.

These $Z$-pole observables have been measured with good precision.
The experimental and SM values of $\Gamma_{Z\to l^+l^-}$ are~\cite{pdg}, in MeV,
\begin{eqnarray} \label{xGZ2ll} &
\Gamma_{Z\to e^+e^-}^{\rm exp} \,\,=\,\, 83.91 \pm 0.12 ~, \hspace{5ex}
\Gamma_{Z\to\mu^+\mu^-}^{\rm exp} \,\,=\,\, 83.99 \pm 0.18 ~, \hspace{5ex}
\Gamma_{Z\to\tau^+\tau^-}^{\rm exp} \,\,=\,\, 84.08 \pm 0.22 ~, & \nonumber \\ &
\Gamma_{Z\to e^+e^-}^{\rm SM} \,\,=\,\, \Gamma_{Z\to\mu^+\mu^-}^{\rm SM} \,\,=\,\,
84.01 \pm 0.07 ~, \hspace{5ex}
\Gamma_{Z\to\tau^+\tau^-}^{\rm SM} \,\,=\,\, 83.82\pm 0.07 ~,
\end{eqnarray}
while those of $A_l^{}$ are~\cite{pdg}
\begin{eqnarray} \label{xAl} &
A_e^{\rm exp} \,\,=\,\, 0.1515\pm0.0019 ~, \hspace{5ex}
A_\mu^{\rm exp} \,\,=\,\, 0.142\pm 0.015 ~, \hspace{5ex}
A_\tau^{\rm exp} \,\,=\,\, 0.143 \pm 0.004 ~, & \nonumber \\ &
A_e^{\rm SM} \,\,=\,\, A_\mu^{\rm SM} \,\,=\,\, A_\tau^{\rm SM} \,\,=\,\, 0.1475 \pm 0.0010 ~. &
\end{eqnarray}
For \,$Z\to\nu\bar\nu$,\, we have~\cite{pdg}, also in MeV,
\begin{eqnarray} \label{zGZ2nn}
\Gamma_{Z\to\rm invisible}^{\rm exp} \,\,=\,\, 499.0\pm1.5 ~, \hspace{5ex}
\Gamma_{Z\to\rm invisible}^{\rm sm} \,\,=\,\, 501.69 \pm 0.06 ~.
\end{eqnarray}

Numerically, for the SM contributions we employ the tree-level formulas in
Eqs. (\ref{MZ2ll})-(\ref{gsm}) and (\ref{MZ2nn}) along with the effective values
\begin{eqnarray} \label{eff}
g_{\rm eff}^{} \,\,=\,\, 0.6517 ~, \hspace{5ex} s_{\rm w,eff}^2 \,\,=\,\, 0.23146
\end{eqnarray}
which allow us to reproduce the SM numbers in Eqs.~(\ref{xGZ2ll}) and (\ref{xAl}) within
their errors and obtain \,$\Gamma_{Z\to\nu\bar\nu}^{\rm sm}=501.26$\,MeV\, in agreement
with~$\Gamma_{Z\to\rm invisible}^{\rm sm}$, indicating that other SM invisible modes are
negligible.
To extract the upper limit on $\bigl|b_{\sf C}^{ll}\bigr|$, we assume it to be the only
nonvanishing coupling subject to the 90\% confidence-level (CL) ranges
\begin{eqnarray} \label{zpole} &
83.71 \,\le\, \Gamma_{Z\to e^+e^-}^{} \,\le\, 84.11 \,, \hspace{5ex}
83.69 \,\le\, \Gamma_{Z\to\mu^+\mu^-}^{} \,\le\, 84.29 \,, \hspace{5ex}
83.72 \,\le\, \Gamma_{Z\to\tau^+\tau^-}^{} \,\le\, 84.44 \,, &  \nonumber \\ &
0.1459 \,\le\, A_e^{} \,\le\, 0.1546 \,, \hspace{5ex}
0.117 \,\le\, A_\mu^{} \,\le\, 0.167 \,, \hspace{5ex}
0.136 \,\le\, A_\tau^{} \,\le\, 0.150 \,, & \nonumber \\ &
497 \,\,\le\,\, \Gamma_{Z\to\nu\bar\nu}^{} \,\,\le\,\, 502 ~,
\end{eqnarray}
the rate numbers being in MeV.\footnote{We have taken the lower (upper) bound of $A_e$
$(\Gamma_{Z\to\nu\bar\nu})$ to be its SM lower (upper) value because $A_e^{\rm exp}$ is
above $A_e^{\rm sm}$ $\bigl(\Gamma_{Z\to\rm invisible}^{\rm exp}$ is below
$\Gamma_{Z\to\rm invisible}^{\rm sm}\bigr)$ by {\small$\sim$\,}2 sigmas.}

\begin{figure}[t]
\includegraphics[width=85mm]{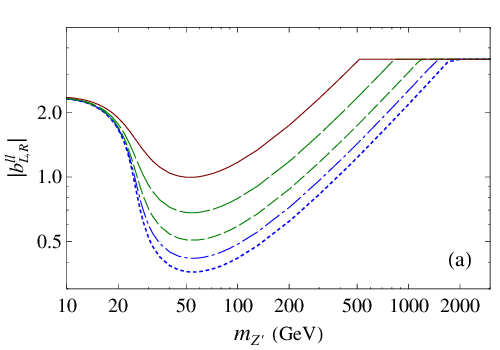}
\includegraphics[width=90mm]{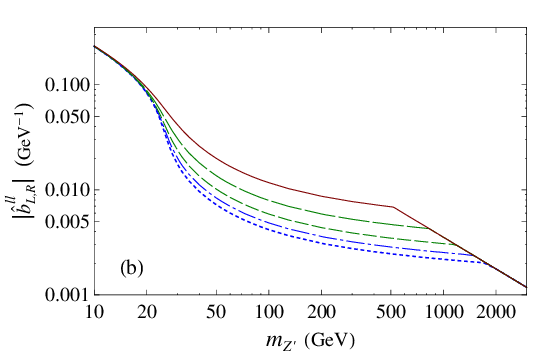} \vspace{-1ex}
\caption{Upper limits on (a)~$|b_{L,R}^{ll}|$  and
(b)~$\bigl|\hat b_{L,R}^{ll}\bigr|=|b_{L,R}^{ll}|/m_{Z'}^{}$\, for \,$l=e,\mu,\tau$\,
at 90\%$\;$CL from $Z$ decay data if only one of the couplings is not zero at a time.
The blue dotted (dotted-dashed) curve refers to~$b_{L(R)}^{ee}$, the green short-dashed
(long-dashed) curve $b_{L(R)}^{\mu\mu}$, and the red solid curve~$b_R^{\tau\tau}$.
The curves for $b_L^{\tau\tau}$ and $b_L^{\mu\mu}$ coincide due to the \,$Z\to\nu\bar\nu$\,
constraint.
The straight portions of the solid curves denote the perturbativity requirement,
\,$|b_{L,R}^{ll}|<\sqrt{4\pi}$.\,
These results also serve as limits on the flavor-changing couplings $|b_{L,R}^{ll'}|$, as
explained in the text.\label{bll-limits}}
\end{figure}

We find that the constraint on $b_L^{ll}$ from \,$Z\to\nu\bar\nu$\, is weaker (stronger) than
that from \,$Z\to l^+l^-$\, for \, $l=e$\, $(l=\mu,\tau)$.
Assuming that only one flavor-conserving coupling is nonzero at a time, we show the results
for \,$10{\rm\,GeV}\le m_{Z'}^{}\le3$\,TeV\, in Fig.\,\ref{bll-limits}(a), where the displayed
curves represent the stronger limit in each $l$ case.
The blue dotted (dotted-dashed) curve refers to $b_{L(R)}^{ee}$,
the green short-dashed (long-dashed) curve $b_{L(R)}^{\mu\mu}$,
and the red solid curve $b_R^{\tau\tau}$.
The curve for $b_L^{\tau\tau}$ coincides with that for $b_L^{\mu\mu}$ because of
the \,$Z\to\nu\bar\nu$\, constraint.
The horizontal solid straight line for $m_{Z'}^{}$ above 500\,GeV marks the perturbativity
limit, \,$|b_{L,R}^{ll}|<\sqrt{4\pi}$,\, which we have imposed as an extra requirement on
the couplings.
We present another view on the results in Fig.\,\ref{bll-limits}(b), which has the corresponding
limits on the couplings divided by the $Z'$ mass.

If instead only one of the flavor-changing couplings is nonvanishing at a time, we can get
its upper limit also from~Fig.\,\ref{bll-limits}, in view of $\epsilon_{\sf C}^{llZ}$
in~Eq.\,(\ref{epsilonllZ}).
Accordingly, the limits on $|b_{\sf C}^{e\mu,\mu e}|$, $|b_{\sf C}^{e\tau,\tau e}|$, and
$|b_{\sf C}^{\mu\tau,\tau\mu}|$ are the same as those on $|b_{\sf C}^{ee}|$, $|b_{\sf C}^{ee}|$,
and $|b_{\sf C}^{\mu \mu}|$, respectively, for \,${\sf C}=L$ or~$R$.\,
As we will see later, there may be stronger bounds on some of these individual
flavor-changing couplings from other measurements, depending on the $Z'$ mass.

Since $Z'$-mediated diagrams can contribute at tree level to \,$e^+e^-\to l^+l^-$\, scattering,
its data can provide additional restrictions on the $Z'$ couplings.
Here we will use LEP-II measurements at various center-of-mass energies above
the $Z$ pole, from 130 to~207~GeV\,~\cite{Alcaraz:2006mx}.
In the absence of $Z$-$Z'$ mixing, the amplitude if \,$l\neq e$\, is
\begin{eqnarray} \label{Mee2ll}
{\cal M}_{\bar e e\to\bar l l}^{} &=&
-4\pi\,\alpha\; \frac{\bar l\gamma^\rho l\, \bar e\gamma_\rho^{}e}{s} \,-\,
\frac{\bar l\gamma^\rho\bigl(g_L^{\rm sm}P_L^{}+g_R^{\rm sm}P_R^{}\bigr)l\,
\bar e\gamma_\rho^{}\bigl(g_L^{\rm sm}P_L^{}+g_R^{\rm sm}P_R^{}\bigr)e}
{s-m_Z^2+i\Gamma_Z^{}m_Z^{}}
\nonumber \\ && -\;
\frac{\bar l\gamma^\rho\bigl(b_L^{ll}P_L^{}+b_R^{ll}P_R^{}\bigr)l\,
\bar e\gamma_\rho^{}\bigl(b_L^{ee}P_L^{}+b_R^{ee}P_R^{}\bigr)e}
{s-m_{Z'}^2+i\Gamma_{Z'}^{}m_{Z'}^{}}
\,+\,
\frac{\bar l\gamma^\rho\bigl(b_L^{le}P_L^{}+b_R^{le}P_R^{}\bigr)e\,
\bar e\gamma_\rho^{}\bigl(b_L^{el}P_L^{}+b_R^{el}P_R^{}\bigr)l}{t-m_{Z'}^2} ~,
\nonumber \\
\end{eqnarray}
where $\alpha$ is the electromagnetic fine-structure constant, the lepton masses have been
neglected, $\Gamma_{Z,Z'}$ denote the total widths, the plus sign of the $t$-channel term
follows from Fermi statistics, \,$s=\bigl(p_{e^+}^{}+p_{e^-}^{}\bigr){}^2$,\, and
\,$t=\bigl(p_{e^-}^{}-p_{l^-}^{}\bigr){}^2$.\,

In the absence of flavor-changing couplings, thus the last line of Eq.\,(\ref{Mee2ll}),
the resulting cross-section \,$\sigma_{e\bar e\to l\bar l}^{}$\, and forward-backward asymmetry
\,$A_{\rm FB}=\sigma_{e\bar e\to l\bar l}^{\rm FB}/\sigma_{e\bar e\to l\bar l}^{}$,\, with
\,$\sigma_{e\bar e\to l\bar l}^{\rm FB}=
\sigma_{e\bar e\to l\bar l}^{\rm F}-\sigma_{e\bar e\to l\bar l}^{\rm B}$,\,
are known in the literature (see, {\it e.g.}, \cite{Chiang:2011cv,Kors:2005uz}).
As mentioned in Ref.\,\cite{Chiang:2011cv} (which also has the more general formulas in
the presence of $Z$-$Z'$ mixing),
the expressions for these observables imply that in a model-independent study their experimental
values cannot lead to restrictions on the individual flavor-conserving $Z'$ couplings, assumed
to be free parameters, but can nevertheless still restrain their
products,~\,$b_{\sf C}^{ee}b_{\sf C'}^{ll}$.

\begin{figure}[b]
\includegraphics[width=84.5mm]{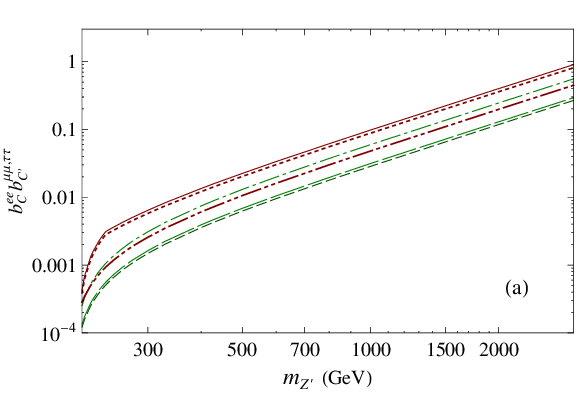} ~
\includegraphics[width=88mm]{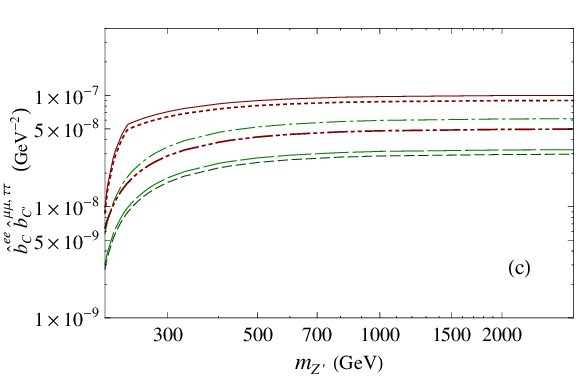} \vspace{-1ex} \\
\includegraphics[width=84.5mm]{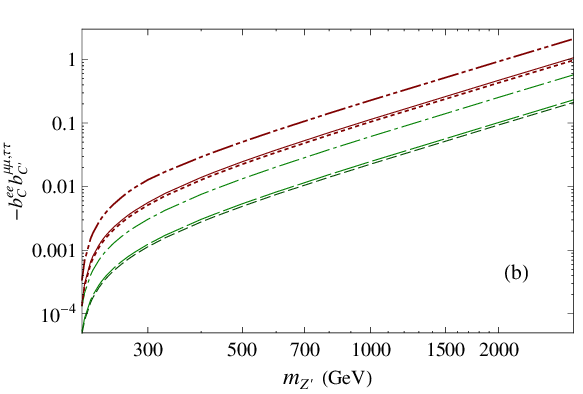} ~
\includegraphics[width=88mm]{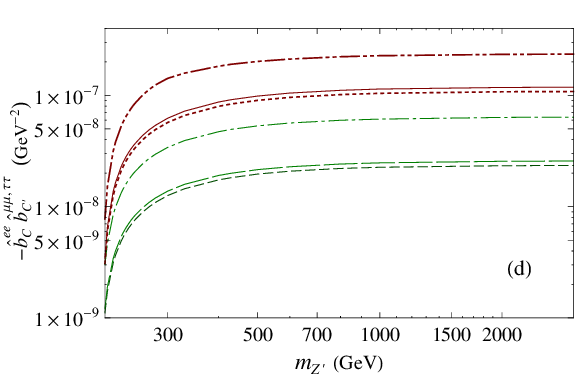} \vspace{-1ex}
\caption{Upper limits on (a)~$b_{\sf C}^{ee}b_{\sf C'}^{ll}$ and
(b)~$-b_{\sf C}^{ee}b_{\sf C'}^{ll}$ for \,$l=\mu,\tau$\, and \,${\sf C,C}'=L,R$\, at 90\%$\;$CL
from LEP\,II data if only one of the coupling products is nonvanishing at a time.
The different curves are described in the text.
The right plots (c and d) show the corresponding limits on
\,$\pm\hat b_{\sf C}^{ee}\hat b_{\sf C'}^{ll}=\pm b_{\sf C}^{ee}b_{\sf C'}^{ll}/m_{Z'}^2$.
\label{beebll-limits}}
\end{figure}

Since we have left the $Z'$ total width, $\Gamma_{Z'}$, unspecified in concentrating on
its couplings to leptons and since it would be needed to compute the cross sections
if~\,$s\sim m_{Z'}^2$,\, we evaluate the limits on \,$b_{\sf C}^{ee}b_{\sf C'}^{ll}$\, from
the LEP\,II data only for $m_{Z'}^{}$ values starting from 210\,GeV up to~3\,TeV.
Using as before~Eq.\,(\ref{eff}), along with the effective value
\,$\alpha_{\rm eff}^{}=1/132.4$,\, and the 90\%$\;$CL ranges of
the experimental numbers~\cite{Alcaraz:2006mx},\footnote{A few of
the $\sigma_{e\bar e\to l\bar l}$ and $A_{\rm FB}$ measurements disagree with their
SM predictions by about $2\;$sigmas or more. In each of those cases, we take the lower (upper)
bound of the required range of the relevant observable to be its SM lower (upper) value if
the measurement is above (below) the SM prediction.} we draw
Fig.\,\ref{beebll-limits}(a) for the upper limits on the products \,$b_L^{ee}b_L^{\mu\mu}$
(green long-dashed curve), $b_{L,R}^{ee}b_{R,L}^{\mu\mu}$ (green dotted-dashed curve),
$b_R^{ee}b_R^{\mu\mu}$ (green short-dashed curve), $b_L^{ee}b_L^{\tau\tau}$ (red dotted curve),
$b_{L,R}^{ee}b_{R,L}^{\tau\tau}$ (red double-dotted-dashed curve), and
$b_R^{ee}b_R^{\tau\tau}$ (red solid curve).\,
Since the positive and negative limits on these coupling products are not generally symmetric
with respect to zero, we present Fig.\,\ref{beebll-limits}(b) for the negative limits.
The plots on the right (c and d) depict the corresponding limits for
\,$\pm\hat b_{\sf C}^{ee}\hat b_{\sf C'}^{ll}=\pm b_{\sf C}^{ee}b_{\sf C'}^{ll}/m_{Z'}^2$.\,

\begin{figure}[hb]
\includegraphics[width=83mm]{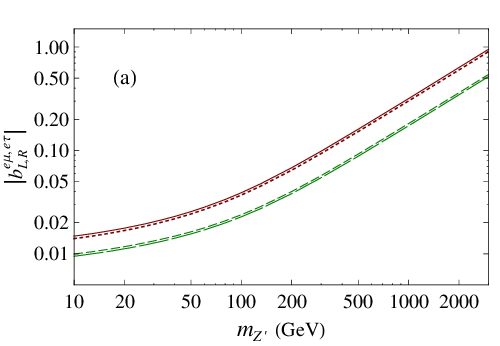} ~
\includegraphics[width=89mm]{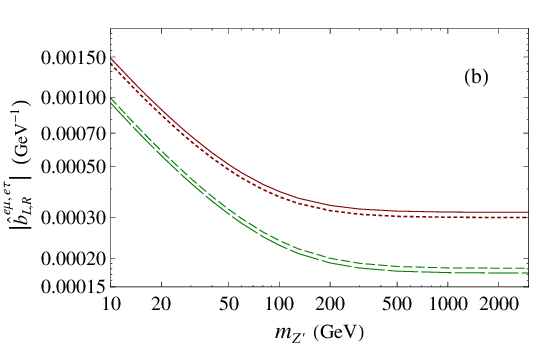} \vspace{-1ex}
\caption{Upper limits on (a)~$|b_{L,R}^{el}|$ and
(b)~$\bigl|\hat b_{L,R}^{el}\bigr|=|b_{L,R}^{el}|/m_{Z'}^{}$
for \,$l=\mu,\tau$\, at 90\%$\;$CL from LEP\,II data.
The green long-dashed (short-dashed) curves refer to $\bigl|b_{L(R)}^{e\mu}\bigr|$
and the red dotted (solid) curves $\bigl|b_{L(R)}^{e\tau}\bigr|$.
\label{bel-limits}}
\end{figure}

If one of the flavor-changing couplings \,$b_{L,R}^{e\mu,e\tau}$\, in Eq.~(\ref{Mee2ll})
does not vanish, we can constrain it separately, assuming \,$b_{L,R}^{ee,ll}=0$.\,
In that case, the \,$e^+e^-\to l^+l^-$\, cross-sections have the expressions given in
the Appendix, and so $\Gamma_{Z'}$ is not required in the calculation.
Using the LEP\,II data again, we obtain the upper limits on $|b_{L,R}^{el}|$ for
\,$10{\rm\,GeV}\le m_{Z'}^{}\le3$\,TeV\, in Fig.\,\ref{bel-limits}.
The second plot displays the corresponding limits on
\,$\bigl|\hat b_{\sf C}^{el}\bigr|=|b_{\sf C}^{el}|/m_{Z'}^{}$.\,
These results are evidently more stringent than the bounds on $|b_{L,R}^{e\mu,e\tau}|$
inferred from Fig.~\ref{bll-limits}.

\subsection{Low-energy processes\label{low}}

Turning our attention now to constraints on the $Z'$ couplings from low-energy data, we will
consider a number of lepton flavor-violating processes and leptonic anomalous magnetic moments.
The relevant formulas are available from the expressions given in
Ref.~\cite{Chiang:2011cv} in the more general case with $Z$-$Z'$ mixing.

We begin by mentioning an additional process that can restrict $b_{L,R}^{e\mu}$ separately,
namely the muonium-antimuonium conversion, \,$\mu^+e^-\to\mu^-e^+$.\,
In this case, the constraints are the same as those determined in~Ref.~\cite{Chiang:2011cv},
\begin{eqnarray} \label{gc}
\bigl|\hat b_{L,R}^{e\mu}\bigr| \,\,\le\,\, 4.4\times10^{-4}{\rm~GeV}^{-1} ~.
\end{eqnarray}
These are stricter than their counterparts in Fig.\,\ref{bel-limits} if $m_{Z'}$ goes below~30\,GeV.

We next look at various constraints on the products of a pair of different couplings of the~$Z'$,
at least one of them being flavor changing, from the experimental limits~\cite{pdg} for
flavor-changing leptonic 3-body and radiative 2-body decays of the $\mu$ and $\tau$ leptons.
The leptonic decays arise from tree-level $Z'$-mediated diagrams, whereas the radiative
decays proceed from loop diagrams containing the $Z'$ and an internal lepton.
Assuming that only the two couplings in each of the products are present at a time, we collect
the results in Table$\;$\ref{le-limits}.

The anomalous magnetic moments $a_e^{}$ and $a_\mu^{}$ of the electron and muon have been
measured very precisely and therefore can provide extra constraints.
In contrast, the experimental information on $a_\tau^{}$ is still too limited to provide
significant bounds~\cite{pdg}.
For the $Z'$ contributions to $a_e^{}$ and $a_\mu^{}$, we will retain only the terms induced
by the $\tau$ lepton in the loop, as they are enhanced by the $\tau$ mass
compared to the other lepton terms~\cite{Chiang:2011cv},
\begin{eqnarray} \label{al}
a_e^{Z'} \,\,=\,\, \frac{m_e^{}m_\tau^{}\,{\rm Re}\bigl(b_L^{e\tau}b_R^{\tau e}\bigr)}
{4\pi^2\,m_{Z'}^2} ~, \hspace{5ex}
a_\mu^{Z'} \,\,=\,\,
\frac{m_\mu^{} m_\tau^{}\,{\rm Re}\bigl(b_L^{\mu\tau}b_R^{\tau\mu}\bigr)}{4\pi^2\,m_{Z'}^2} ~.
\end{eqnarray}
The SM prediction for $a_e^{}$ is compatible with its most recent measurement, the difference
between them being \,$a_e^{\rm exp}-a_e^{\rm SM}=(-105\pm 82)\times10^{-14}$~\cite{Aoyama:2012wj}.
Consequently, we can impose the 90\%$\;$CL
range~\,$-2.4\times10^{-12}\le a_e^{Z'}\le0.3\times10^{-12}$.\,
In contrast, the SM and experimental values of $a_\mu^{}$ presently differ by
nearly 3 sigmas,
\,$a_\mu^{\rm exp}-a_\mu^{\rm SM}=(249\pm87)\times10^{-11}$~\cite{Aoyama:2012wk}.
This suggests that we may require \,$0\le a_\mu^{Z'}\le3.3\times10^{-9}$.\,
It follows that
\begin{eqnarray} \label{g-2}
-1.0\times10^{-7} \,\le\, {\rm Re}\bigl(\hat b_L^{e\tau}\hat b_R^{\tau e}\bigr) {\rm\,GeV}^2
\,\le\, 0.1\times10^{-7} ~, \hspace{5ex}
\color{black}
0 \,\le\, {\rm Re}\bigl(\hat b_L^{\mu\tau\,}\hat b_R^{\tau\mu}\bigr){\rm\,GeV}^2
\,\le\, 6.9\times10^{-7} ~. ~~~~
\end{eqnarray}
The $\hat b_L^{e\tau}\hat b_R^{\tau e}$ limits are complementary to the individual bounds on
$\hat b_{L,R}^{e\tau}$ illustrated in Fig.\,\ref{bel-limits}.

\begin{table}[h]
\caption{Limits on products of $Z'$ couplings determined from low-energy data.
The second column contains 90\%$\;$CL experimental upper-limits on the branching ratios of
the listed decay modes.\label{le-limits}}
\small
\begin{tabular}{|c|c|c|} \hline
~~ Decay ~~ & Measured &
Derived limits on products of \, $\hat b=b/m_{Z'}^{}\vphantom{\int^|}$ \\
mode & limits~\cite{pdg} & in \,GeV$^{-2}$ \\
\hline\hline
$\mu\to e e\bar e\vphantom{\int_{|_|}^\int}$ & $1.0\times10^{-12}$ &
$\bigl| \hat b_{\sf C}^{ee\,}\hat b_{\sf C}^{e\mu} \bigr| \,\le\, 2.3\times10^{-11} ~, ~~~~
\bigl| \hat b_{L,R\,}^{ee}\hat b_{R,L}^{e\mu} \bigr| \,\le\, 3.3\times10^{-11}$ \\
$\tau\to ee\bar e\vphantom{\int_{|_|}^|}$       & $2.7\times10^{-8}$ &
$\bigl| \hat b_{\sf C}^{ee\,}\hat b_{\sf C}^{e\tau} \bigr| \,\le\, 9.1\times10^{-9} ~, ~~~~
\bigl| \hat b_{L,R\,}^{ee}\hat b_{R,L}^{e\tau} \bigr| \,\le\, 1.3\times10^{-8}$ \\
$\tau\to\mu\mu\bar\mu\vphantom{\int_{|_|}^|}$  & $2.1\times10^{-8}$ &
$\bigl| \hat b_{\sf C}^{\mu\mu\,}\hat b_{\sf C}^{\mu\tau} \bigr| \,\le\, 8.0\times10^{-9} ~, ~~~~
\bigl| \hat b_{L,R\,}^{\mu\mu}\hat b_{R,L}^{\mu\tau} \bigr| \,\le\, 1.1\times10^{-8}$ \\
$\tau\to\mu e\bar e\vphantom{\int_{|_|}^|}$      & $1.8\times10^{-8}$ &
$\bigl| \hat b_{\sf C}^{ee\,}\hat b_{\sf C'}^{\mu\tau} \bigr| \,\le\, 1.0\times10^{-8} ~, ~~~~
\bigl| \hat b_{\sf C}^{e\mu\,}\hat b_{\sf C'}^{e\tau} \bigr| \,\le\, 1.0\times10^{-8}$ \\
$\tau\to e\mu\bar\mu\vphantom{\int_{|_|}^|}$   & $2.7\times10^{-8}$ &
$\bigl| \hat b_{\sf C}^{\mu\mu\,}\hat b_{\sf C'}^{e\tau} \bigr| \,\le\, 1.3\times10^{-8} ~, ~~~~
\bigl| \hat b_{\sf C}^{e\mu\,}\hat b_{\sf C'}^{\mu\tau} \bigr| \,\le\, 1.3\times10^{-8}$ \\
$\tau\to ee\bar\mu\vphantom{\int_{|_|}^|}$    & $1.5\times10^{-8}$ &
$\bigl| \hat b_{\sf C}^{e\mu\,}\hat b_{\sf C}^{e\tau} \bigr| \,\le\, 6.8\times10^{-9} ~, ~~~~
\bigl| \hat b_{L,R\,}^{e\mu}\hat b_{R,L}^{e\tau} \bigr| \,\le\, 9.6\times10^{-9}$ \\
$\tau\to\mu\mu\bar e\vphantom{\int_{|_|}^|}$   & $1.7\times10^{-8}$ &
$\bigl| \hat b_{\sf C}^{e\mu\,}\hat b_{\sf C}^{\mu\tau} \bigr| \,\le\, 7.2\times10^{-9} ~, ~~~~
\bigl| \hat b_{L,R\,}^{e\mu}\hat b_{R,L}^{\mu\tau} \bigr| \,\le\, 1.0\times10^{-8}$
\\
$\mu\to e\gamma\vphantom{\int_{|_|}^|}$      & \,$2.4\times10^{-12}$\, &
\,$\bigl| \hat b_{\sf C}^{e\mu\,}\hat b_{\sf C}^{\mu\mu} \bigr| \le 1.3\times10^{-9} \,, ~~
\bigl| \hat b_{L,R\,}^{e\mu}\hat b_{R,L}^{\mu\mu} \bigr| \le 4.3\times10^{-10} \,, ~~
\bigl| \hat b_{L,R\,}^{e\tau}\hat b_{R,L}^{\tau\mu} \bigr| \le 2.6\times10^{-11}$\, \\
$\tau\to e\gamma\vphantom{\int_{|_|}^|}$     & $3.3\times10^{-8}$ &
$\bigl| \hat b_{\sf C}^{ee,\tau\tau\,}\hat b_{\sf C}^{e\tau} \bigr| \le 3.6\times10^{-7} \,, ~~
\bigl| \hat b_{\sf C}^{e\mu\,}\hat b_{\sf C}^{\mu\tau} \bigr| \le 3.6\times10^{-7} \,, ~~
\bigl| \hat b_{L,R\,}^{e\tau}\hat b_{R,L}^{\tau\tau} \bigr| \le 1.2\times10^{-7}$ \\
$\tau\to\mu\gamma\vphantom{\int_{|_|}^|}$    & $4.4\times10^{-8}$ &
$\bigl| \hat b_{\sf C}^{\mu\mu,\tau\tau\,}\hat b_{\sf C}^{\mu\tau} \bigr| \le 4.2\times10^{-7} \,, ~~
\bigl| \hat b_{\sf C}^{\mu e\,}\hat b_{\sf C}^{e\tau} \bigr| \le 4.2\times10^{-7} \,, ~~
\bigl| \hat b_{L,R\,}^{\mu\tau}\hat b_{R,L}^{\tau\tau} \bigr| \le 1.4\times10^{-7}$ \\
\hline
\end{tabular}
\end{table}

\section{Predictions\label{sec:predictions}}

The results above allow us to explore how the $Z'$ effects may modify the leptonic decays of
the newly found particle, $h$, assumed to be a SM-like Higgs boson, and also those of
the weak bosons~($Z$ and~$W$).
We will deal with both flavor-conserving and -violating channels involving two and four
leptons in the final states.

\subsection{Two-body decays\label{2body}}

We first look at the flavor-violating decay \,$h\to l^\pm l^{\prime\mp}$.\,
It proceeds from a $Z'$-loop diagram having an $hl^+l^-$ and two lepton-$Z'$ vertices,
at least one of the latter being flavor changing, and one-loop \,$l\to l'$\, diagrams with
flavor-changing couplings.
We calculate the amplitude to be
\begin{eqnarray} \label{h2ll'}
{\cal M}_{h\to l^{\prime+}l^-}^{} \,\,=\,\, \frac{\sqrt{m_l^{}m_{l'}^{}}}{v}\;
\bar l \bigl( \epsilon_L^{ll'h}P_L^{}+\epsilon_R^{ll'h}P_R^{} \bigr) l' ~,
\end{eqnarray}
where \,$v=246$\,GeV\, is the Higgs vacuum expectation value,
\begin{eqnarray} \label{ell'h} & \displaystyle
\epsilon_{L,R}^{ll'h} \,\,=\,\, {\cal F}_h(r)\,\mbox{\footnotesize$\displaystyle\sum_f$}\,
\frac{m_f^{}}{\sqrt{m_l^{}m_{l'}^{}}}\,b_{R,L\,}^{lf}b_{L,R}^{fl'} ~, \hspace{7ex}
r \,\,=\,\, \frac{m_h^2}{m_{Z'}^2} ~, & ~~~~~~~
\nonumber \\ & \displaystyle
{\cal F}_h(r) \,\,=\,\, \frac{-1}{8\pi^2}
\bigl[ \ln r\,\ln(1+r) + {\rm Li}_2(-r) \,-\, i\pi\, \ln(1+r) \bigr] ~, &
\end{eqnarray}
with $f$ being the fermion in the loop.
The rate for \,$m_{l,l'}^2\ll m_h^2$\, is then
\begin{eqnarray}
\Gamma_{h\to l^{\prime+}l^-}^{} \,\,=\,\, \frac{m_{h\,}^{}m_{l\,}^{}m_{l'}^{}}{16\pi\,v^2}
\Bigl( \bigl|\epsilon_L^{ll'h}\bigr|^2+\bigl|\epsilon_R^{ll'h}\bigr|^2\Bigr) ~.
\end{eqnarray}

To predict the largest rates, we observe from the bounds derived in the last section and
Eq.\,(\ref{ell'h}) that the most important contribution comes from the internal lepton
\,$f=\tau$\, and that the rates are maximized for final states with one $\tau$.
Thus, we take
\,$|b_{L,R\,}^{\tau\tau}b_{R,L}^{\tau e}|=1.2\times10^{-7}\,m_{Z'}^2/{\rm GeV}^2$\, and
\,$|b_{L,R\,}^{\tau\tau}b_{R,L}^{\tau\mu}|=1.4\times10^{-7}\,m_{Z'}^2/{\rm GeV}^2$\,
from the \,$\tau\to e\gamma,\mu\gamma$\, bounds in Table$\;$\ref{le-limits}.
For definiteness, we take \,$m_h^{}=125.5$\,GeV,\, compatible with the average
$h$ mass of \,$125.7\pm0.4$~GeV\, from the LHC measurements~\cite{lhc}.
With only one nonzero product of couplings being present in each case
and~\,$\Gamma_h=4.14$\,MeV\,~\cite{lhctwiki},
we  obtain for~\,$10{\rm\,GeV}\le m_{Z'}^{}\le3$\,TeV\,
\begin{eqnarray}
{\cal B}(h\to\mu\tau) \,\,=\,\, {\cal B}(h\to\mu^+\tau^-)+{\cal B}(h\to\mu^-\tau^+)
\,\,\lesssim\,\, 3\times10^{-9} ~,
\end{eqnarray}
the upper bound occurring at \,$m_{Z'}^{}=3$\,TeV,\, and a somewhat smaller number
for ${\cal B}(h\to e\tau)$.
Clearly these $Z'$-induced flavor-violating Higgs decays will not be observable in the near future.

We next consider the $Z'$ impact on the flavor-conserving decay \,$h\to l^+l^-$.\,
The $Z'$ contribution follows from Eq.\,(\ref{h2ll'}) after setting \,$l'=l$.\,
Combining the result with the SM tree-level contribution, we then get
\begin{eqnarray}
{\cal M}_{h\to l^+l^-}^{} \,\,=\,\, \frac{m_l}{v}\, \bar l\,\Bigl[
\Bigl(1+\epsilon_L^{llh}\Bigr) P_L^{}+\Bigl(1+\epsilon_R^{llh}\Bigr)P_R^{}\Bigr] l
\end{eqnarray}
leading to
\begin{eqnarray}
\Gamma_{h\to l^+l^-}^{} \,\,=\,\, \frac{m_{h\,}^{}m_l^2}{16\pi\,v^2}\Bigl(
\bigl|1+\epsilon_L^{llh}\bigr|^2+\bigl|1+\epsilon_R^{llh}\bigr|^2\Bigr) ~.
\end{eqnarray}

\begin{figure}[b]
\includegraphics[height=68mm]{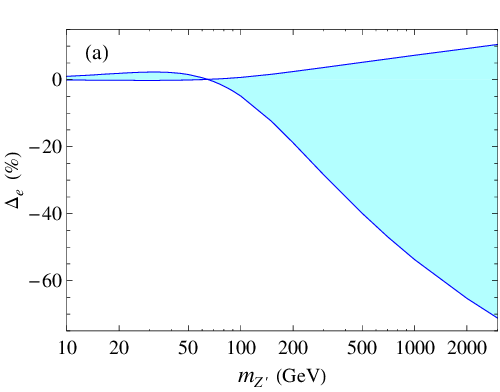} ~
\includegraphics[height=68mm]{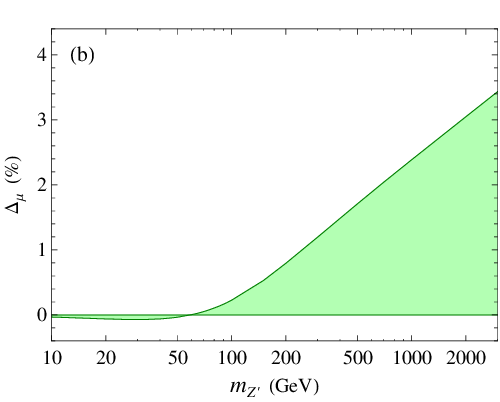} \vspace{-1ex}
\caption{Range of fractional change to SM rate of \,$h\to l^+l^-$\, for (a)~$l=e$ and
(b)~$l=\mu$ due to only the coupling products $b_{R,L}^{e\tau\,}b_{L,R}^{\tau e}$ and
$b_{R,L}^{\mu\tau\,}b_{L,R}^{\tau\mu}$, respectively.\label{deltaem}}
\end{figure}

We expect again that the internal lepton \,$f=\tau$\, in the loop yields the maximal impact.
Accordingly, for \,$h\to e^+e^-$\, and \,$h\to\mu^+\mu^-$\, we take, respectively,
the $b_{R,L}^{e\tau\,}b_{L,R}^{\tau e}$ and \,$b_{R,L}^{\mu\tau\,}b_{L,R}^{\tau\mu}$\, ranges
in~Eq.\,(\ref{g-2}), assuming that the coupling products are purely real or dominated by
the real part.
The graphs in Fig.\,\ref{deltaem} depict the resulting fractional change
\begin{eqnarray}
\Delta_l^{} \,\,=\,\, \frac{\Gamma_{h\to l^+l^-}^{}}{\Gamma_{h\to l^+l^-}^{\rm sm}} \,-\, 1
\end{eqnarray}
in the rate due to the presence of $Z'$ exclusively via these flavor-changing couplings.
Although the $Z'$ contribution can reduce the \,$h\to e^+e^-$\, rate sizeably
[blue-shaded areas in Fig.\,\ref{deltaem}(a)], this decay mode,
with a branching ratio of \,$\mbox{\footnotesize$\sim$}\,5\times10^{-9}$\, in the SM,
may be beyond reach for a long time.
Much more interesting is \,$h\to\mu^+\mu^-$,\, which has a SM branching ratio of
\,$\mbox{\footnotesize$\sim$}\,2\times10^{-4}$\,
and therefore may be measurable in the not-so-distant future with precision possibly
sensitive to the $Z'$ effect, indicated by the green-shaded areas in~Fig.\,\ref{deltaem}(b).

\begin{figure}[t]
\includegraphics[height=67mm]{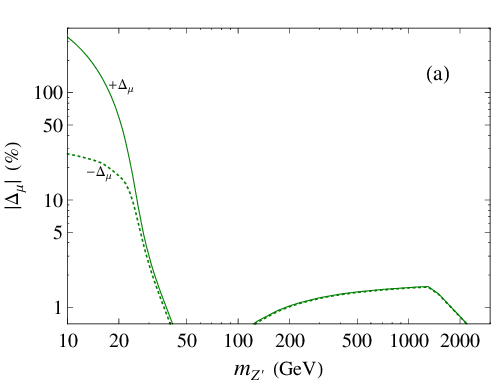} \,
\includegraphics[height=67mm]{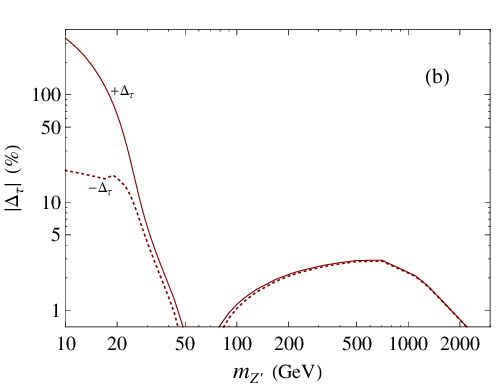} \vspace{-1ex}
\caption{Upper limits on fractional changes $\Delta_l$ and $-\Delta_l$ to SM rate of
\,$h\to l^+l^-$\, for (a)~$l=\mu$ and (b)~$l=\tau$ if only
the flavor-conserving couplings $b_{L,R}^{ll}$ are nonzero.\label{deltamt}}
\end{figure}

As it turns out, potentially more considerable modifications to
the \,$h\to l^+l^-$\, rate can be induced by
the flavor-conserving couplings $b_{L,R}^{ll}$, which enter $\epsilon_{L,R}^{llh}$
as the product $b_L^{ll}b_R^{ll}$.
To estimate their maximal impact, we focus on the \,$l=\mu,\tau$\, cases.
In each of them, we set all the other couplings to zero and scan the values of $b_{L,R}^{ll}$
satisfying the requirements in~Eq.\,(\ref{zpole}) as well as the perturbativity
condition~\,$|b_{L,R}^{ll}|<\sqrt{4\pi}$.\,
We find that the coupling values allowed by these constraints can translate into substantial
$\Delta_{\mu,\tau}$ that are positive or negative.
Especially, in the \,$10{\rm\,GeV}\le m_{Z'}^{}<50$\,GeV\, region, the decrease in the rate
could reach a few tens percent, whereas the increase could exceed~100\%, even up to
{\small$\sim$\,}300\%, as can be seen in~Fig.\,\ref{deltamt}.

The \,$h\to\tau^+\tau^-$\, decay has begun to be observable at the LHC.
The latest signal strengths for this channel reported by the ATLAS and CMS Collaborations
are \,$\sigma/\sigma_{\rm sm}^{}=0.8\pm0.7$\, and \,$0.7\pm0.5$,\, respectively, at
\,$m_h^{}=125$\,GeV~\cite{h2tt}.
Obviously, these early findings already disfavor some parts of the $Z'$ parameter space implied
by~Fig.\,\ref{deltamt}(b), although it is still too soon to be quantitative about the exclusion
zones in view of the sizable uncertainties of these current data.
As their precision continues to improve, the upcoming experiments can either uncover a~$Z'$
signal or restrain its $\tau$ couplings further.
A similar expectation can be stated regarding the $\mu$ couplings from future measurements of
the \,$h\to\mu^+\mu^-$\, mode.

The flavor-violating decays \,$Z\to\bar l'l$\, are not yet observed, but have been searched
for, the experimental upper-limits on their branching ratios being~\cite{pdg}
\begin{eqnarray}
{\cal B}(Z\to e\mu) \,\le\, 1.7\times10^{-6} \,, ~~~~
{\cal B}(Z\to e\tau) \,\le\, 9.8\times10^{-6} \,, ~~~~
{\cal B}(Z\to\mu\tau) \,\le\, 1.2\times10^{-5} \,, ~~~
\end{eqnarray}
at 95\%$\;$CL, where
\,${\cal B}(Z\to l_1 l_2)={\cal B}\bigl(Z\to l_1^+l_2^-\bigr) +
{\cal B}\bigl(Z\to l_1^-l_2^+\bigr)$.\,
These modes get $Z'$-loop contributions analogously to those in \,$Z\to l^+l^-$\,
and hence may provide further limits on the $Z'$ couplings if the experimental
limits are less than the predicted values derived from the upper limits on the couplings.
To make the predictions for the $e\mu$, $e\tau$, and $\mu\tau$ final states,
we take the biggest values
\begin{eqnarray}
\label{biggest}
\bigl|\hat b_{\sf C}^{e\mu\,}\hat b_{\sf C}^{\mu\mu}\bigr| \,\,=\,\, 1.3\times10^{-9} ~,
\hspace{5ex}
\bigl|\hat b_{\sf C}^{e\tau\,}\hat b_{\sf C}^{\tau\tau}\bigr| \,\,=\,\, 3.6\times10^{-7} ~,
\hspace{5ex}
\bigl|\hat b_{\sf C}^{\mu\tau\,}\hat b_{\sf C}^{\tau\tau}\bigr| \,\,=\,\, 4.2\times10^{-7}
\end{eqnarray}
from the \,$\mu\to e\gamma$,\, \,$\tau\to e\gamma$,\, and \,$\tau\to\mu\gamma$\, bounds,
respectively, in Table$\;$\ref{le-limits}.
These translate into
\begin{eqnarray}
{\cal B}(Z\to e\mu)_{Z'} \,<\, 7.1\times10^{-12} \,, ~~~
{\cal B}(Z\to e\tau)_{Z'} \,<\, 5.4\times10^{-7} \,, ~~~
{\cal B}(Z\to\mu\tau)_{Z'} \,<\, 7.4\times10^{-7} \,. ~~~
\end{eqnarray}
Hence, although the $e\mu$ mode is likely to be undetectable, the $e\tau$ and $\mu\tau$
predictions are only 18 and 16 times below their respective experimental bounds.
It is worth noting that these results are comparable to those found in~Ref.\,\cite{Chiang:2011cv}
in the presence of $Z$-$Z'$ mixing, in accord with the expectation mentioned therein.

\subsection{Four-body decays\label{4body}}

The presence of the $Z'$ may also manifest itself in the decays of $h$ and the weak bosons
into four leptons.
The CMS Collaboration~\cite{CMS:2012bw} has recently reported the first observation of $Z$
decays into four charged leptons ($e$ and $\mu$) consistent with the SM expectations.
This suggests that such decays can be measured well in the future at the LHC, thereby
providing another means to look for new-physics hints, such as those of the $Z'$.
On the other hand, since four-lepton final states in $h$ decays proceed mainly from the \,$h\to ZZ^*$\,
or \,$h\to WW^*$\, mode, the $Z'$ impact would expectedly be very small and hard to detect in
flavor-conserving decays.
The flavor-violating four-lepton decays of $h$ would also have rates that are too tiny
to be observable.
We therefore focus on the $Z'$ contributions to a number of 4-lepton decays of the gauge bosons,
\,$Z\to l_1^+l_2^-l_3^+l_4^-$\, and \,$W^{+(-)}\to l_1^+l_2^-l_3^+\nu\bigl(l_3^-\bar{\nu}\bigr)$\,
for \,$l=e,\mu,\tau$,\, using the constraints found in Section~\ref{sec:constraints}.
We compute these decays using the CalcHEP~3.4 package~\cite{Ref:CalcHEP} incorporating the new
vertices in the model file and assuming \,$m_{Z'}^{}\geq 210$\,GeV,\, since we have not specified
the $Z'$ total width.

In the flavor-conserving decays \,$Z\to l_1^+l_1^-l_2^+l_2^-$,\,
the $Z'$ contributions depend on the $b_{\sf C}^{ll}$ couplings.
Applying the constraints given in Fig.\,\ref{bll-limits}, we estimate the possible ranges of
the fractional change
\,$\Delta^Z_{4 l}=\Gamma_{Z \to l^+l^-l^+l^-}/\Gamma_{Z \to l^+l^-l^+l^-}^{\rm sm}-1$\,
for \,$l_1=l_2=l=e,\mu,\tau$,\, taking only one of $b_{\sf C}^{ll}$ to be nonzero at a~time.
Thus, we obtain
\begin{eqnarray}
-0.0032 \,\,\leq\,\, \Delta^Z_{4 e} \,\,\leq\,\, 0 ~, \hspace{5ex}
-0.016 \,\,\leq\,\, \Delta^Z_{4 \mu} \,\,\leq\,\,  0 ~, \hspace{5ex}
-0.033 \,\,\leq\,\,  \Delta^Z_{4 \tau} \,\,\leq\,\, 0 ~,
\end{eqnarray}
where the lowest numbers all belong to \,$m_{Z'}^{}=210$\,GeV\, and
we have imposed the kinematical cut \,$M_{\bar{l}l} > 4$\,GeV\, in estimating the rates.
Since the SM predicts
\,${\cal B}(Z \to l^+ l^- l^+ l^-)_{\rm sm}\simeq 1.2\times 10^{-6}$\,
for~\,$l=e,\mu$~\cite{CMS:2012bw}, a large luminosity is required to observe the changes.
We also compute the possible ranges of the fractional change
\,$\Delta^Z_{2 l2 l'}=
\Gamma_{Z\to l^+l^-l^{\prime+}l^{\prime-}}/\Gamma_{Z\to l^+l^-l^{\prime+}l^{\prime-}}^{\rm sm}-1$\,
in the \,$l\neq l'$\, cases where~\,$l^{(\prime)}=e, \mu, \tau$.\,
For \,$(l,l')=(e, \mu)$ and $(e, \tau)$\, we should take into account
the constraints on the products of a pair of different couplings in Fig.\,\ref{beebll-limits},
whereas for \,$(l,l')=(\mu, \tau)$\, we apply the upper limits on $|b^{\mu \mu}_{\sf C}|$ and
$|b^{\tau \tau}_{\sf C}|$ from Fig.\,\ref{bll-limits}.
We find that the magnitudes of $\Delta^Z_{2e 2\mu}$ and $\Delta^Z_{2e 2\tau}$ are smaller than
${\cal O}\bigl(10^{-3}\bigr)$ because of the constraints on $|b_{\sf C}^{ee} b_{\sf C}^{\mu \mu}|$
and $|b_{\sf C}^{ee} b_{\sf C}^{\tau \tau}|$ in Fig.\,\ref{beebll-limits}, while
the magnitude of $\Delta^Z_{2\mu 2 \tau}$ can be comparatively greater,
\begin{eqnarray}
-0.013 \,\,\leq\,\, \Delta^Z_{2 \mu 2 \tau} \,\,\leq\,\, 0.014 ~,
\end{eqnarray}
the positive (negative) sign of $\Delta^Z_{2 \mu 2 \tau}$ corresponding to
the negative (positive) sign of $b_{ R}^{\mu \mu} b_{ R}^{\tau \tau}$.
One can see that the upper bound of $|\Delta^Z_{2\mu 2 \tau}|$ is roughly similar
in order of magnitude to the lower bounds of $\Delta^Z_{4\mu}$ and $\Delta^Z_{4 \tau}$.

We also consider the flavor-violating 4-lepton decays of the $Z$,
concentrating on the modes \,$Z \to \tau^+ \tau^- \tau^\pm e^\mp $,\,
\,$Z \to\tau^+\tau^-\tau^\pm\mu^\mp $,\, and \,$Z \to \tau^\pm\tau^\pm\mu^\mp\mu^\mp $,\,
as their amplitudes are determined by
$\bigl|\hat b_{\sf C}^{e \tau\,}\hat b_{\sf C}^{\tau\tau}\bigr|$,
$\bigl|\hat b_{\sf C}^{\mu \tau\,}\hat b_{\sf C}^{\tau\tau}\bigr|$, and
$|\hat b_{\sf C}^{\mu \tau}|^2$, respectively.
The constraints on the first two products of couplings are given in Table$\;$\ref{le-limits},
and the upper limits on $|b_{\sf C}^{\mu \tau}|$ come from Fig.\,\ref{bll-limits}.
Applying the biggest values from Eq.\,(\ref{biggest}) and the upper limit on
$|b_{ L}^{\mu \tau}|$ in Fig.\,\ref{bll-limits}, we find that the branching ratios
are of ${\cal O}\bigl(10^{-13}\bigr)$ for the first two modes and ${\cal O}\bigl(10^{-9}\bigr)$
for the third one, which will be unobservable in the near future.

The 4-lepton decays of the $W$ are treated similarly.
In the flavor-conserving case, the fractional change
\,$\Delta^W_{3 l}=\Gamma_{W^+\to l^+l^-l^+\nu}/\Gamma_{W^+\to l^+l^-l^+\nu}^{\rm sm}-1$\,
for \,$l=e,\mu,\tau$\, depend on $|b_{L}^{ll}|$, in light of
the relation between $b_{\nu}^{ll}$ and $b_{L}^{ll}$ in Eq.~(\ref{bb*}).
We obtain
\begin{eqnarray}
-0.0083 \,\,\leq\,\, \Delta^W_{3 e} \,\,\leq\,\, 0 ~,  \hspace{5ex}
-0.012 \,\,\leq\,\, \Delta^W_{3 \mu, 3\tau} \,\,\leq\,\, 0 ~,
\end{eqnarray}
where the lowest value corresponds to \,$m_{Z'}= 210$\,GeV\, and we have employed
the \,$M_{\bar{l}l} > 4$\,GeV\, cut as before in estimating the rates.
The $\Delta^W_{3 \mu, 3\tau}$ numbers are almost the same because the upper limits of
$|b_L^{\mu\mu,\tau\tau}|$ are the same.
Since the SM prediction is
\,${\cal B}\bigl(W^{+(-)}\to l^+ l^- l^+\nu(l^-\bar\nu)\bigr) \simeq 1.1 \times 10^{-6}$,\,
again a large luminosity is required to measure the changes, as in the $Z$-decay case.

For the flavor-violating 4-lepton decays of the $W$, we consider the
\,$W^{+(-)}\to \tau^\pm \tau^\pm e^-\nu(e^+\bar\nu)$\, and
\,$W^{+(-)}\to \tau^\pm \tau^\pm \mu^-\nu(\mu^+\bar\nu)$\, channels.
We obtain the biggest branching ratios of these modes from the upper limits of
$\bigl|\hat b_{L}^{e \tau\,}\hat b_{L}^{\tau\tau}\bigr|$ and $|b_{ L}^{\mu \tau}|$, respectively.
Using~Eq.\,(\ref{biggest}) and Fig.\,\ref{bll-limits}, we find that the branching ratio of
the first mode is at most of ${\cal O}\bigl(10^{-12}\bigr)$ and that of the second mode
${\cal O}\bigl(10^{-9}\bigr)$.
Hence they will likely be undetectable for a long time, as the flavor-violating 4-lepton $Z$ decays.

\section{Conclusions \label{sec:summary}}

We have explored a $Z'$ boson originating from a new U(1) gauge symmetry and interacting with
leptons in a family-nonuniversal way.  The $Z'$ is assumed to have no mixing with the $Z$ boson.
We have studied the effects of the $Z'$ as a virtual particle in various processes with leptons
in the initial and/or final states, especially the leptonic decays of the newly discovered $h$,
putatively a SM-like Higgs boson.

We first study bounds on the leptonic couplings of the $Z'$ from available experimental data.
We employ the $Z$-pole observables, including the \,$Z \to l^+ l^-$\, rates, the associated
forward-backward asymmetries, and the invisible $Z$ decay rate, to put constraints
on $|b_{L,R}^{ll}|$.
The cross sections of \,$e^+e^- \to l \bar l$\, for \,$l = \mu,\tau$\, from LEP-II experiments
are used to place bounds on the products of chiral couplings $b^{ee}_{\sf C} b^{ll}_{\sf C}$
for \,${\sf C} = L,R$.\,  Further restrictions are found from various low-energy processes,
including the muonium-antimuonium conversion for $|b_{L,R}^{e\mu}|$ in particular, several
flavor-changing leptonic 3-body and radiative 2-body decays of the muon and tauon, and
the anomalous magnetic moments of the electron and muon.

We then apply the constraints on the $Z'$ couplings and mass to make predictions for
the flavor-conserving and -violating 2-body leptonic decays of $h$ as well as 4-body decays of
the $W$ and $Z$.  For the flavor-violating \,$h \to l' \bar l$\, decays, which arise from
one-loop diagrams, the most important contribution comes from the case with one $\tau$ in
the final state and an internal $\tau$ running in the loop due to mass enhancement.
Unfortunately, such $Z'$-mediated flavor-violating Higgs decays have rates that are too small
to be observable in the near future.  On the other hand, the flavor-conserving \,$h\to l^+l^-$\,
decay can be significantly affected by the contribution of an internal~$\tau$.
With flavor-changing couplings alone, the branching ratio of \,$h \to \mu^+\mu^-$\, decay may
be enhanced by a mere few percent.  In contrast, with only flavor-conserving couplings, both
the \,$h \to \mu^+\mu^-$\, and \,$h \to \tau^+\tau^-$\, channels can be considerably modified
by a few tens to a few hundreds percent.
Consequently, upcoming measurements of the latter mode with better precision than its current
data will either uncover or constrain the $Z'$ boson.  Additional tests can be expected from
future findings on \,$h\to\mu^+\mu^-$\, at the LHC or a Higgs factory.

We have found that the flavor-violating \,$Z \to\bar l'l$\, decays, mediated by the $Z'$ at
the loop level, are consistent with current experimental upper bounds at 95\% CL.
The 4-body leptonic decays of the $Z$ have started to be measured by the LHC.
Our calculations show that a larger luminosity is required to observe modifications in
flavor-conserving \,$Z \to 4l$\, decays due to the $Z'$, the maximal changes being
of~${\cal O}(10^{-2})$.
A similar situation holds for flavor-conserving \,$W \to 3l+\nu$\, decays.
We have also found that the $Z'$-mediated flavor-violating \,$Z \to 4l$\, and
\,$W \to 3l+\nu$\, decays are unlikely to be observable soon.

\acknowledgments

We would like to thank Xiao-Gang He and German Valencia for helpful discussions.
This research was supported in part by the National Science Council of Taiwan, R.O.C.,
under Grant No.~NSC-100-2628-M-008-003-MY4.

\appendix

\section{Cross sections of \,$\bm{e^+e^-\to l^+l^-}$\label{csafb}}

If the flavor-changing couplings in Eq.\,(\ref{Mee2ll}) are absent, the cross sections are
known in the literature~\cite{Chiang:2011cv,Kors:2005uz}.
If instead \,$b_{L,R}^{ee,ll}=0$,\, one derives from Eq.\,(\ref{Mee2ll})
\begin{eqnarray} \label{csee2ll}
\sigma_{e\bar e\to l\bar l}^{} &=& \frac{4\pi\alpha^2}{3\,s} \,+\,
\frac{\alpha}{6}\; \frac{\bigl(g_L^{\rm sm}+g_R^{\rm sm}\bigr)\raisebox{1pt}{$^2$}\bigl(s-m_Z^2\bigr)}
{\bigl(s-m_Z^2\bigr)\raisebox{1pt}{$^2$}+\Gamma_Z^2m_Z^2}
\,+\,
\frac{\bigl[ (g_L^{\rm sm})^2+(g_R^{\rm sm})^2 \bigr]\raisebox{1pt}{$^2$}s}
{48\pi\,\bigl[\bigl(s-m_Z^2\bigr)\raisebox{1pt}{$^2$}+\Gamma_Z^2m_Z^2\bigr]}
\nonumber \\ && +\;
\Biggl\{ \frac{\alpha\,\bigl[\bigl(b_L^{el}\bigr)\raisebox{1pt}{$^2$}
+ \bigl(b_R^{el}\bigr)\raisebox{1pt}{$^2$}\bigr]}{4\,s^3}
+ \frac{\bigl[\bigl(g_L^{\rm sm}\bigr)\raisebox{1pt}{$^2$}\bigl(b_L^{el}\bigr)\raisebox{1pt}{$^2$}
+ \bigl(g_R^{\rm sm}\bigr)\raisebox{1pt}{$^2$}\bigl(b_R^{el}\bigr)\raisebox{1pt}{$^2$}\bigr]
\bigl(s-m_Z^2\bigr)}
{16\pi\,\bigl[\bigl(s-m_Z^2\bigr)\raisebox{1pt}{$^2$}+\Gamma_Z^2m_Z^2\bigr]s^2} \Biggr\}
\nonumber \\ && ~\times
\Biggl[ 2m_{Z'}^2s+3s^2+2\bigl(m_{Z'}^2+s\bigr)\raisebox{1pt}{$^2$}\,\ln\frac{m_{Z'}^2}{m_{Z'}^2+s} \Biggr]
\nonumber \\ && +\;
\frac{\bigl(b_L^{el}\bigr)\raisebox{1pt}{$^4$}+\bigl(b_R^{el}\bigr)\raisebox{1pt}{$^4$}}{16\pi\,m_{Z'}^2s^2}
\Biggl[ 2m_{Z'}^2s+s^2+2m_{Z'}^2\bigl(m_{Z'}^2+s\bigr)\,\ln\frac{m_{Z'}^2}{m_{Z'}^2+s} \Biggr]
\,+\,
\frac{\bigl(b_L^{el}\bigr){}^2\,\bigl(b_R^{el}\bigr)\raisebox{1pt}{$^2$}\,s}
{8\pi\,m_{Z'}^2\bigl(m_{Z'}^2+s\bigr)} ~, ~~~~~~~
\end{eqnarray}
\begin{eqnarray} \label{csfbee2ll}
\sigma_{e\bar e\to l\bar l}^{\rm FB} &=&
\frac{\alpha}{8}\; 
\frac{\bigl(g_L^{\rm sm}-g_R^{\rm sm}\bigr)\raisebox{1pt}{$^2$}\bigl(s-m_Z^2\bigr)}
{\bigl(s-m_Z^2\bigr)\raisebox{1pt}{$^2$}+\Gamma_Z^2m_Z^2}
\,+\,
\frac{\bigl[\bigl(g_L^{\rm sm}\bigr)\raisebox{1pt}{$^2$} 
- \bigl(g_R^{\rm sm}\bigr)\raisebox{1pt}{$^2$}\bigr]\raisebox{1pt}{$^2$}s}
{64\pi\,\bigl[\bigl(s-m_Z^2\bigr)\raisebox{1pt}{$^2$}+\Gamma_Z^2m_Z^2\bigr]}
\nonumber \\ && +\;
\Biggl\{ \frac{\alpha\,\bigl[\bigl(b_L^{el}\bigr)\raisebox{1pt}{$^2$}
+ \bigl(b_R^{el}\bigr)\raisebox{1pt}{$^2$}\bigr]}{8\,s^3} +
\frac{\bigl[\bigl(g_L^{\rm sm}\bigr)\raisebox{1pt}{$^2$}\bigl(b_L^{el}\bigr)\raisebox{1pt}{$^2$} + 
\bigl(g_R^{\rm sm}\bigr)\raisebox{1pt}{$^2$}\bigl(b_R^{el}\bigr)\raisebox{1pt}{$^2$}\bigr]
\bigl(s-m_Z^2\bigr)}
{32\pi\,\bigl[\bigl(s-m_Z^2\bigr)\raisebox{1pt}{$^2$}+\Gamma_Z^2m_Z^2\bigr]s^2} \Biggr\}
\nonumber \\ && ~\times
\Biggl[ s^2 + 4\bigl(m_{Z'}^2+s\bigr)\raisebox{1pt}{$^2$}\,
\ln\frac{4m_{Z'}^2\bigl(m_{Z'}^2+s\bigr)}{\bigl(2m_{Z'}^2+s\bigr)\raisebox{1pt}{$^2$}} \Biggr]
\nonumber \\ && +\;
\frac{\bigl[\bigl(b_L^{el}\bigr)\raisebox{1pt}{$^4$} + 
\bigl(b_R^{el}\bigr)\raisebox{1pt}{$^4$}\bigr]\bigl(m_{Z'}^2+s\bigr)}
{16\pi\,m_{Z'}^2\bigl(2m_{Z'}^2+s\bigr)s^2}
\Biggl[ s^2 + 2m_{Z'}^2\bigl(2m_{Z'}^2+s\bigr)\,
\ln\frac{4m_{Z'}^2\bigl(m_{Z'}^2+s\bigr)}{\bigl(2m_{Z'}^2+s\bigr)\raisebox{1pt}{$^2$}} \Biggr]
\nonumber \\ && +\;
\frac{\bigl(b_L^{el}\bigr)\raisebox{1pt}{$^2$}\,\bigl(b_R^{el}\bigr)\raisebox{1pt}{$^2$}\,s^2}
{8\pi\,m_{Z'}^2\bigl(m_{Z'}^2+s\bigr)\bigl(2m_{Z'}^2+s\bigr)} ~, ~~~~
\end{eqnarray}
where we have assumed $g_{L,R}^{el}$ to be real and neglected the lepton masses.

\end{document}